\documentclass[graphicx,superscriptaddress,longbibliography,reprint]{revtex4-1}

\usepackage[version=3]{mhchem} % Formula subscripts using \ce{}
\usepackage[T1]{fontenc}
\usepackage{siunitx}
\usepackage{graphicx}

\usepackage[normalem]{ulem}
\useunder{\uline}{\ul}{}
\begin{document}
\author{Robin J. Dolleman}
\affiliation{Kavli Institute of Nanoscience, Delft University of Technology, Lorentzweg 1, 2628 CJ, Delft, The Netherlands}
\email{R.J.Dolleman@tudelft.nl}
\author{Dejan Davidovikj}
\affiliation{Kavli Institute of Nanoscience, Delft University of Technology, Lorentzweg 1, 2628 CJ, Delft, The Netherlands}
\author{Herre S. J. van der Zant}
\affiliation{Kavli Institute of Nanoscience, Delft University of Technology, Lorentzweg 1, 2628 CJ, Delft, The Netherlands}
\author{Peter G. Steeneken}
\affiliation{Kavli Institute of Nanoscience, Delft University of Technology, Lorentzweg 1, 2628 CJ, Delft, The Netherlands}
\affiliation{Department of Precision and Microsystems Engineering, Delft University of Technology, Mekelweg 2, 2628 CD, Delft, The Netherlands}
\title{Amplitude calibration of 2D mechanical resonators by nonlinear optical transduction}

\begin{abstract}
Contactless characterization of mechanical resonances using Fabry-Perot interferometry is a powerful tool to study the mechanical and dynamical properties of atomically thin membranes.  However, amplitude calibration is often not performed, or only possible by making assumptions on the device parameters such as its mass or the temperature. In this work, we demonstrate a calibration technique that directly measures the oscillation amplitude by detecting higher harmonics that arise from nonlinearities in the optical transduction. Employing this technique, we calibrate the resonance amplitude of two-dimensional nanomechanical resonators, without requiring knowledge of their mechanical properties, actuation force, geometric distances or the laser intensity.
\end{abstract}
\maketitle

%\section*{Introduction}
The dynamics of 2D material resonators has spurred an enormous interest because of their sensitivity to the surrounding environment, paving the way towards gas \cite{dolleman2016graphene,koenig2012selective} and pressure sensors \cite{dolleman2015graphene,bunch2008impermeable}. Additionally, the intricate thermal \cite{dolleman2017optomechanics}, optical \cite{barton2012photothermal} and mechanical properties \cite{davidovikj2017young} of these materials are of interest as well. The analysis of the linear frequency response of suspended 2D membranes usually provides information on their pre-tension $n_0$ through the resonance frequency $f_0$ and on their energy dissipation rate through the quality factor $Q$. Besides $f_0$ and $Q$, it is often desirable to calibrate the amplitude of the resonant motion. This enables force sensing and also allows for determination of the mass, Young's modulus \cite{davidovikj2017young} and the thermal properties \cite{dolleman2017optomechanics}. However, current calibration techniques assume that the temperature or the mass are well known, which is difficult to justify for 2D material membranes. 
 
Readout of the dynamic displacement of 2D resonators is usually performed by the following two methods: (i) transconductance measurements \cite{zande2010large,chen2016modulation,chen2009performance,singh2012coupling}, where motion is detected via a gate-induced conductance modulation or (ii) laser interferometry \cite{davidovikj2016visualizing,castellanos2013single,bunch2007electromechanical,barton2011high,zande2010large,lee2013high}, where a Fabry-Perot cavity is formed between the resonator and a fixed mirror so that the motion of the resonator modulates the intensity of the reflected light. Thermomechanical calibration of the amplitude relies on the equipartition theorem \cite{hauer2013general}. This method is widely used for calibrating cantilevers for atomic force microscopy \cite{hauer2013general} and has recently been applied to few-layer graphene resonators \cite{davidovikj2016visualizing,davidovikj2017young}. When applied to single-layer 2D materials however, thermomechanical calibration has the drawback that one has to assume that both the temperature and modal mass are known. The mass can be significantly affected by impurities and polymer contamination \cite{chen2009performance}, therefore resulting in considerable errors in the calibration of the motion amplitude of the membrane.
 
At high amplitudes the assumption of a linear transduction coefficient breaks down, since the output signal is no longer proportional to the displacement. In Fabry-Perot interferometry this happens because the intensity of the reflected light is a periodic function of the membrane's position.  This nonlinear relation between membrane position and the intensity of the reflected light is well known \cite{blake2007making,roddaro2007optical,jung2007simple,casiraghi2007rayleigh,abergel2007visibility,cartamil2016colorimetry,cartamil2017very} and manifests itself in the frequency domain by higher harmonic generation at integer multiples of the driving frequency $f$.  %This nonlinear relation between membrane position and the intensity of the reflected light is well known \cite{blake2007making,roddaro2007optical,jung2007simple,casiraghi2007rayleigh,abergel2007visibility,cartamil2016colorimetry,cartamil2017very}.

Here, we use heterodyne detection to measure these higher harmonics and derive mathematical expressions that relate their intensity ratios to the motion amplitude. We show that using only three harmonics we can deduce both the resonant amplitude and the position of the resonator, i.e. the cavity depth. This procedure provides an alternative for the thermomechanical amplitude calibration method, but is instead independent of the mass and temperature of the resonator and only requires the wavelength of the light to be known.

\begin{figure}
\includegraphics{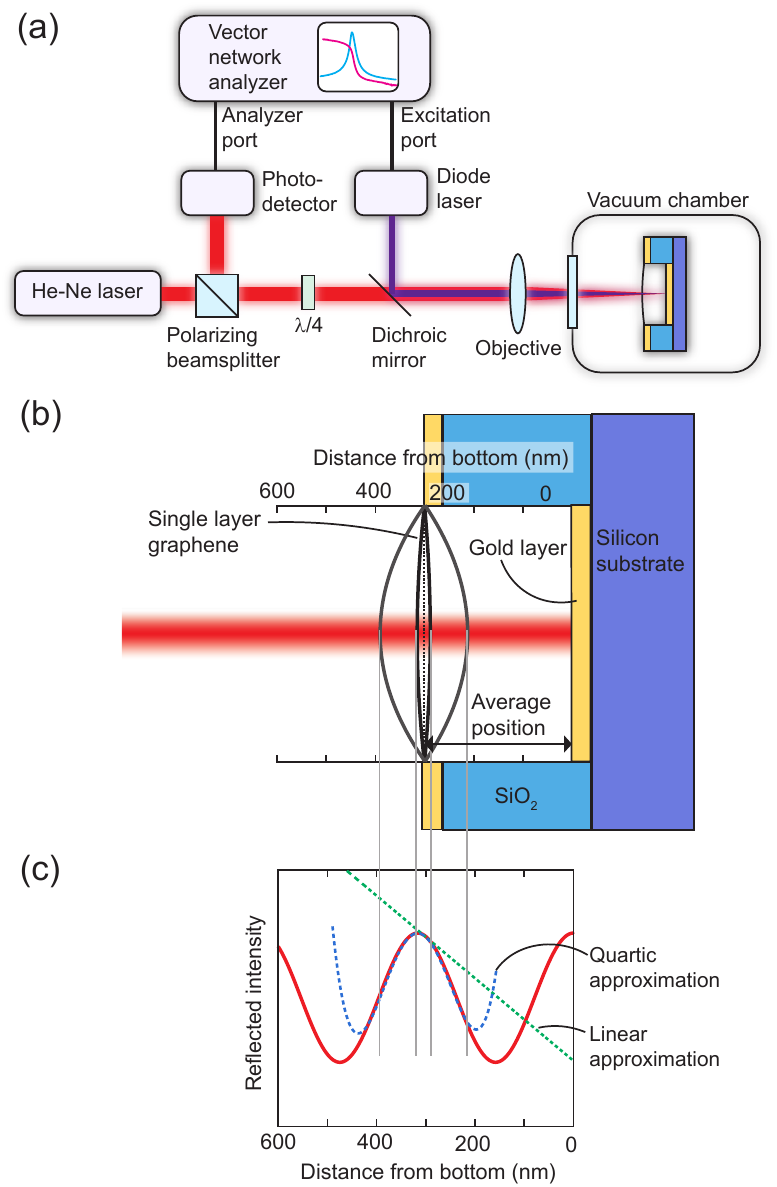}
\caption{(a) Fabry-Perot interferometer setup used in the experiments. (b) Cross section of the suspended graphene device. (c) The reflected intensity detected by the photodetector (solid red line) as a function of membrane distance from the cavity (Eq. \ref{eq:cos}), which deviates from the linear approximation when the amplitude becomes large compared to the wavelength. \label{fig:1}}
 \end{figure}
%\section{Experimental Setup}
We demonstrate the method using a Fabry-Perot interferometer as shown in Fig. \ref{fig:1}(a). A red helium-neon laser with a wavelength of $\lambda =  633$ nm is used for the readout. This laser is focused at the center of a single-layer graphene drum resonator, which is suspended over cavities in a reflective gold substrate (Fig. \ref{fig:1}(b)). These cavities were etched in a layer of 300 nm silicon dioxide, after which a layer of 5 nm chromium and 40 nm gold was evaporated to enhance the optical reflectivity of the substrate. To fabricate graphene drum resonators, a sheet of single layer graphene grown by chemical vapour deposition (CVD) was transfered over the chip. A more detailed description of the samples and their fabrication technique can be found in Ref. \cite{dolleman2017optomechanics}. Due to interference between the moving graphene membrane and the fixed substrate, the reflected intensity of the red laser is a function of the position of the graphene (Fig. \ref{fig:1}(c)). This reflected light is detected by the photodiode. In order to drive the motion of the membrane, a blue diode laser is focused on the resonator. The intensity of this light is modulated, which periodically heats up the membrane slightly and provides a mechanical drive due to thermal expansion. 

An important component in the setup is the vector network analyzer (VNA, type Rohde and Schwarz ZNB4-k4). This apparatus measures the transmission ratio between the modulation voltage of the diode laser and the voltage signal detected at the photodiode. Normally this is done in a homodyne detection scheme, where only the frequency component equal to the driving frequency is detected. However, the frequency conversion option of this VNA enables one to drive the resonator at the resonance frequency, while detecting the photodiode signal at a different frequency. This feature allows detection of the higher harmonics that arise from the nonlinear optical transduction.

We now use optical theory to show how these higher harmonics can be used to determine the motion amplitude and average position. Figure \ref{fig:1}(b) shows a cross-section of the graphene device suspended over the cavity. The reflected intensity of the red laser light (red solid curve in Fig. \ref{fig:1}(c)) is a periodic function of the membrane position, therefore it can be described by a Fourier series. If the membrane is thin enough and the reflectivity of the back mirror is sufficiently high, the reflected intensity $I$ as a function of distance from the cavity bottom can be approximated by a single term in the series (see Supplemental information):
\begin{equation}\label{eq:cos}
I(t) = A + B \cos{\left(4 \pi \frac{g+x(t)}{\lambda}\right)},
\end{equation}
where $A$ and $B$ are constants, $g$ is the average distance between the membrane and the bottom of the cavity, $x$ the membrane's deflection and $\lambda$ the wavelength of the light used for the readout. For small amplitudes a linear approximation can be used for Eq. \ref{eq:cos}, however for large amplitudes this approximation breaks down and a Taylor expansion with more orders is necessary to accurately describe the amplitude (Fig. \ref{fig:1}(c)). Using this Taylor series expansion, it can be mathematically shown that for a sinusoidal motion of the graphene membrane $x(t)= \delta\sin(\omega t)$ the detected optical modulation amplitudes can be expressed by the series $ I(t) = \sum \limits_{m}^{~} I_{m\omega} \sin{m \omega t}$ where $m = 1, 2, 3...$. Performing the series expansion up to $m = 4$ gives for the amplitudes $I_{m\omega}$ (see Supplemental information for the derivation):
 \begin{eqnarray}\label{eq:1om}
 I_{1\omega} =  - B \gamma \delta  \sin{\left(\gamma g \right)} + \frac{1}{8} B \delta^3  \gamma^3 \sin{\left( \gamma g \right)},\\
 I_{2\omega}  =  \frac{1}{4} B \delta^2 \gamma^2  \cos{\left( \gamma g \right)} - \frac{1}{48} B\gamma^4 \delta^4 \cos(\gamma g), \\
 I_{3\omega} = - \frac{1}{24} B \delta^3   \gamma^3 \sin{\left( \gamma g \right)} \label{eq:I3om},  \\
 I_{4\omega}  = \frac{1}{192} B \gamma^4 \delta^4 \cos(\gamma g),
 \end{eqnarray}
where $\gamma = 4 \pi/\lambda$ and higher order terms of $\delta$ are neglected. Note, that $I_{1\omega}$ contains a term linearly proportional to $\delta$, but also a term proportional to $\delta^3$, which causes deviations from linear response in the conventional homodyne Fabry-Perot readout. Using the ratio between the harmonics ${I_{3\omega}}/{I_{1\omega}}$ an expression is obtained that is independent of $A$ and $B$:
 \begin{eqnarray}\label{eq:delta}
 \delta = \frac{ 2\sqrt{6  {I_{3\omega}}/{I_{1\omega}}}}{\sqrt{ \gamma^2-{I_{3\omega}}/{I_{1\omega}}\gamma^2} }.
 \end{eqnarray}
With this equation the amplitude $\delta$ can be determined directly from the measured ratio ${I_{3\omega}}/{I_{1\omega}}$ and the wavelength of the light $\lambda$, since $\gamma = 4 \pi / \lambda$, as is shown in Fig. \ref{fig:calexpl}(a). In the Supplemental information it is shown that the amplitude $\delta$ can also be obtained from the ratio $I_{4\omega}/{I_{2\omega}}$, which can be more accurate when $\sin{(\gamma g)}$ is small. 

\begin{figure}
\includegraphics{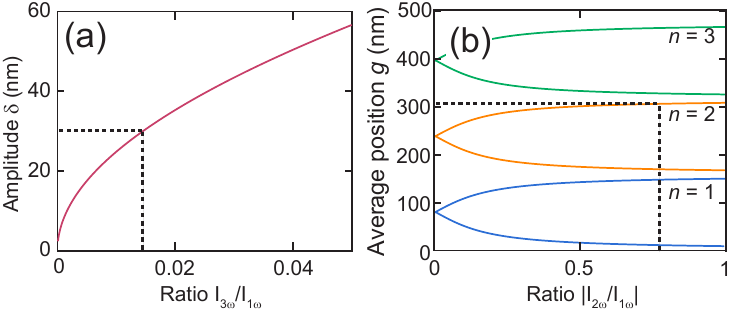}
\caption{Explanation of the calibration procedure. (a) The amplitude $\delta$ of the membrane versus the ratio $I_{3\omega}/I_{1\omega}$. From the measurement of this ratio the amplitude can be directly determined from Eq. \ref{eq:delta}. (b) Average position $g$ versus the ratio $I_{2\omega}/I_{1\omega}$ from Eq. \ref{eq:g}, with a known amplitude of $\delta$. From the measured ratio, the gap size can be determined. However a rough initial guess of this gap size is required to choose the correct value of $n$ in Eq. \ref{eq:g}. \label{fig:calexpl}}
\end{figure}
Once the amplitude $\delta$ is determined from Eq. \ref{eq:delta}, the ratio $I_{2\omega}/ I_{1\omega}$ can now be used to obtain the average position $g$:
\begin{eqnarray}\label{eq:g}
g =  \frac{1}{\gamma} \left(\pi n + \arctan{\left( \frac{12  \delta \gamma- \delta^3 \gamma^3 }{(6 \delta^2  \gamma^2 - 48){I_{2\omega}}/{I_{1\omega}}}\right)} \right)
 \nonumber \\ \mathrm{where } ~ n = 0,1,2,3...
\end{eqnarray}
The procedure to obtain $g$ from this equation is shown in Fig. \ref{fig:calexpl}(b). Note, that the value of $g$ needs to be roughly known from the fabrication process, with an accuracy better than $\lambda/4$, to determine the value of $n$ in Eq. \ref{eq:g}. Since the fabricated depth of the cavities is 300 nm, $n=2$ gives the correct average position in our case (Fig. \ref{fig:calexpl}(b)). It is shown in the Supplemental information that other ratios, such as $I_{3\omega}/{I_{2\omega}}$, yield similar expressions for $g$.

 \begin{figure}
 \includegraphics{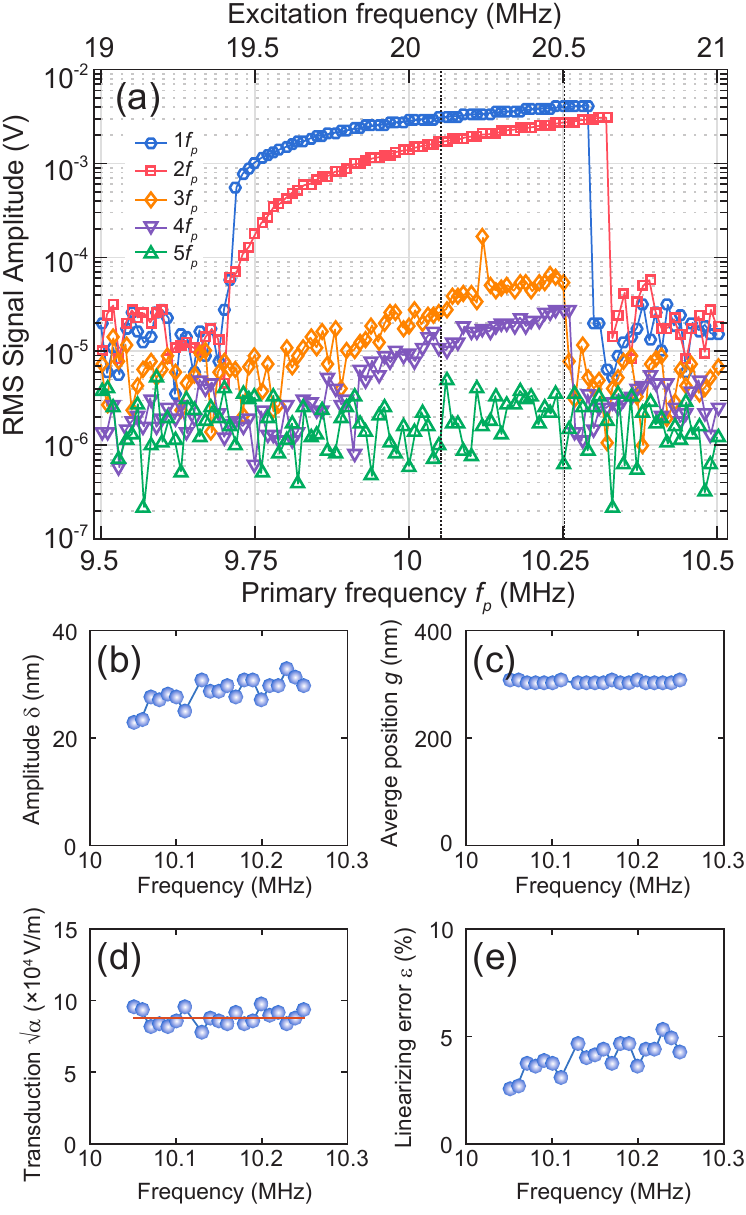}
 \caption{(a) Detection of 5 harmonics of the parametrically driven fundamental mode for a 5 micrometer circular drum. The fifth harmonic has a magnitude smaller than the noise floor; the lower harmonics are readily detected. Dashed lines indicate the window in which the analysis was performed. (b) Amplitude extracted from the data using Eq. \ref{eq:delta}. (c) Average position extracted from the data using Eq. \ref{eq:g}. (d) Transduction coefficient $\sqrt{\alpha}$, the change in root mean square voltage per metre of amplitude of motion. (e) Estimated error in the response by assuming that the transduction is linear. \label{fig:3}}
 \end{figure}
We now experimentally demonstrate the method for a 5 micron diameter, single-layer graphene drum.  Using the setup in Fig. \ref{fig:1} we detect the harmonics due to nonlinear transduction. The intensity modulated laser heats the drum, this causes a tension modulation in the membrane by thermal expansion. Since the spring constant of the membrane is proportional to the tension, this modulation results in a parametric excitation of the drum resonances if the modulation frequency is twice the resonance frequency. Parametric driving was chosen because it resulted in larger amplitudes than direct driving, which increased the accuracy of the calibration method. Parametric excitation was achieved by setting the frequency $f_{\mathrm{ext}}$ of the excitation port of the VNA to twice the primary frequency $f_p$: $f_{\mathrm{ext}} = 2 f_p$. By scanning $f_p$ across the mechanical fundamental resonance frequency $f_0$, the drum is brought into parametric resonance. To detect the first, second, third, fourth and fifth harmonic the frequency of the analyzer port was set to $f_a = f_p$, $2f_p$, $3f_p$, $4f_p$ and $5f_p$ respectively. The resulting signal amplitudes are shown in Fig. \ref{fig:3}(a). In the frequency window indicated by dashed vertical lines in Fig. \ref{fig:3}(a), four harmonics are clearly above the noise level and the calibration procedure can be applied. The data-points are averaged within this frequency window to reduce the error due to measurement noise. 

First, we determine the amplitude of oscillation $\delta$ for all the frequencies in the window using Eq. \ref{eq:delta} (Fig. \ref{fig:3}(b)). A remarkably large amplitude is detected, close to 100 times the thickness of the graphene membrane (0.335 nm), which increases with frequency as expected. Now that the amplitude is known, Eq. \ref{eq:g} is used to find the equilibrium position shown in Fig. \ref{fig:3}(c). An average position of $g = 304.9$ nm is calculated with a standard error (SDE) of 0.16 nm. The transduction coefficient $\sqrt{\alpha}$ is deduces from the relation $I_{1\omega} \approx \sqrt{\alpha} \delta$, by taking the detected root mean square voltage $I_{1\omega}$ at the VNA and dividing it by the amplitude $\delta$ from Fig. \ref{fig:3}(b). The resulting $\sqrt{\alpha} \approx - B \gamma \sin{(\gamma g)}$ within the frequency window is shown in Fig. \ref{fig:3}(d). We find $\sqrt{\alpha} = (\num{8.8}\pm \num{0.1})\times 10^4 $V/m. As expected, the average position $g$ and the transduction coefficient $\sqrt{\alpha}$ are independent of excitation frequency or membrane amplitude. 

\begin{figure}
\includegraphics{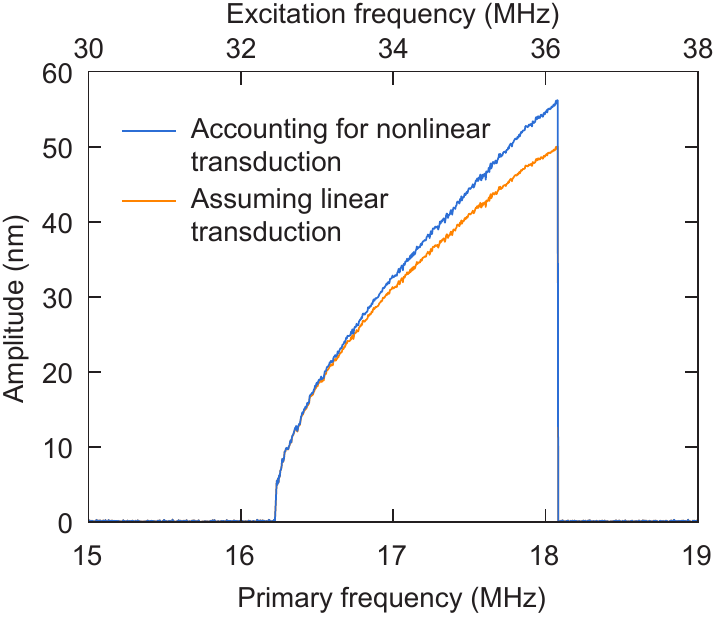} 
\caption{Measured amplitude assuming nonlinear transduction and the corrected signal taking nonlinear transduction into account.\label{fig:4}}
\end{figure}  
The calibration method can also be used to correct for the effects of nonlinear transduction, improving the high-amplitude accuracy of the interferometer. As discussed above, the expression for $I_{1\omega}$ (Eq. \ref{eq:1om}) contains a term proportional to $\delta^3$, which can be used to estimate the relative error $\epsilon$ due to nonlinear transduction, from Eq. \ref{eq:1om}:
 \begin{equation}
\frac{ I_{1\omega}}{\sqrt{\alpha}} = \delta\left( 1 - \frac{1}{8}\delta^2 \gamma^2 \right) \equiv \delta ( 1- \epsilon),
\end{equation}
where $\epsilon = \frac{1}{8} \delta^2  \gamma^2$ \cite{davidovikj2017young}. For small $\epsilon$, the amplitude $\delta$ can now be derived from the uncorrected amplitude $I_{1\omega}/\sqrt{\alpha}$:
\begin{equation}
\delta = (1 + \frac{1}{8} \left(\frac{I_{1\omega}}{\sqrt{\alpha}}\right)^2  \gamma^2) \frac{ I_{1\omega}}{\sqrt{\alpha}},
\end{equation}
with a known value of $\sqrt{\alpha}$ from the calibration, $\delta$ can be found from the measurement of $I_{1\omega}$. Since $\sqrt{\alpha}$ is constant this correction also works outside the frequency window where the calibration is performed. To illustrate the error in the graphene membrane amplitude, we apply this correction to a different drum in Fig. \ref{fig:4}, which exhibits large motion amplitudes. In this case, the maximum amplitude gets underestimated by more than 10\%. This correction is thus important to take into account when measuring the motion of resonators with large amplitudes. 

 The calibration method presented here is demonstrated on single-layer graphene membranes, however it can be extended to other nanomechanical systems, such as nanowires \cite{de2017low}, provided that the assumptions leading to Eq. \ref{eq:cos} are valid. The method could also be extended to thicker membranes, however since Eq. \ref{eq:cos} does not hold anymore in that case, the mathematics become rather complex and may require numerical routines. 
 
 Mechanical nonlinearities have been left out of the analysis, however these could lead to higher harmonics in the mechanical response that could interfere with the calibration. In the Supplemental information, we show that these undesired mechanical nonlinearities can be disentangled from the desired optical nonlinearity by including the fourth harmonic $I_{4\omega}$ in the analysis. This leads to more expressions for $\delta$ and $g$, which are completely independent of mechanical nonlinearities. Extending the analysis can also helps to determine the systemic errors by the simplification behind Eq. \ref{eq:cos}. From the extended analysis, we find that the systemic errors on the transduction coefficient $\sqrt{\alpha}$ are lower than 10\%. This is considerably smaller than existing techniques that require the mass to be known, since the mass can show deviations as high as 600\% \cite{chen2009performance}.
 
   In conclusion, we demonstrate a technique that directly determines the amplitude and average position of suspended single-layer graphene resonators in a Fabry-Perot interferometer. This technique takes advantage of the nonlinear transduction of the membrane motion by detecting the higher harmonics that arise due to optical nonlinearities. The technique can be used to calibrate the motion without any assumptions or knowledge of the mass, the mechanical properties, the actuation force and the intensity of the laser power. Only knowledge of the wavelength of the light is required, thus providing a powerful means towards fully contactless characterization of the mechanical properties of atomically thin membranes.

\begin{acknowledgments}
We gratefully acknowledge Applied Nanolayers B.V. for the growth and transfer of the single-layer graphene used in this study. We further thank Farbod Alijani for discussions. This work is part of the research programme Integrated Graphene Pressure Sensors (IGPS) with project number 13307 which is financed by the Netherlands Organisation for Scientific Research (NWO).
The research leading to these results also received funding from the European Union's Horizon 2020 research and innovation programme under grant agreement No 649953 Graphene Flagship and this work was supported by the Netherlands Organisation for Scientific Research (NWO/OCW), as part of the Frontiers of Nanoscience program.
\end{acknowledgments}

\pagebreak
\onecolumngrid
\setcounter{equation}{0}
\setcounter{figure}{0}
\setcounter{table}{0}
\makeatletter
\renewcommand{\theequation}{S\arabic{equation}}
\renewcommand{\thefigure}{S\arabic{figure}}

\section*{Supplementary Material: Amplitude calibration of 2D mechanical resonators by nonlinear optical transduction }

In this Supplementary section, the analysis of the calibration method is extended. The first section shows how the amplitude can be found using the ratio between $I_{4\omega}/I_{2\omega}$ and the second section shows how mechanical nonlinearities can be disentangled from the optical nonlinearities. These are both applied to our experimental results as presented in the last section. The extended analysis places an upper bound to the systemic errors in the calibration method. For the amplitude, we find that the systemic error is at most 10\% and for the postion detection we find that the systematic error is below 2\%.

\section{Mathematical derivations}
The full expression of the reflected intensity of the laser light $I$ as a function of the membrane's position, $x$, is well known and can be found in various sources \cite{blake2007making,roddaro2007optical,jung2007simple,casiraghi2007rayleigh,abergel2007visibility,cartamil2016colorimetry,cartamil2017very}.
Since this is a periodic function, we express the reflected intensity as a function of deflection as a Fourier series:
\begin{equation}\label{eq:int}
I(x) = \frac{A_0}{2} + \sum_{n=1}^N A_n \sin\left(\frac{4\pi n (g+x)}{\lambda} + \phi_n\right),
\end{equation}
which is periodic in $\lambda/2$. If one assumes that the membrane is optically very thin and the backmirror is an ideal reflector, $I(x)$ can be approximated by the first term in this Fourier series:
\begin{equation}\label{eq:cos}
I(x) =  A + B \cos\left(\frac{4\pi (g+x)}{\lambda}\right).
\end{equation}
This expression can be Taylor expanded; we choose to perform this expansion up to the fourth order since four harmonics are observed in the experiment:
\begin{eqnarray}
I(x) = A + B \cos{\left(\gamma g\right)} - B \gamma \sin{\left(\gamma g \right)}x \nonumber \\ - B \gamma ^2 \cos{\left( \gamma g \right)} \frac{x^2}{2} + B  \gamma^3 \sin{\left( \gamma g \right)} \frac{x^3}{6} + B \gamma^4 \cos(\gamma g) \frac{x^4}{24} + O(5).
\end{eqnarray}
 Now, if it is assumed that the motion of the membrane, $x(t)$, is sinusoidal: $x(t) =  \sin{\omega t}$, the following expression for the time-dependent intensity is obtained:
\begin{eqnarray}
I(t) = \underbrace{A+ B \cos{\left(\gamma g\right)}}_{DC}  \underbrace{- B \gamma \sin{\left(\gamma g \right)}\delta \sin(\omega t)}_{1\omega} \nonumber \\ \underbrace{ - B \gamma^2 \cos{\left( \gamma g \right)} \frac{\delta^2 \sin^2(\omega t)}{2}}_{2\omega} + \underbrace{B  \gamma^3 \sin{\left( \gamma g \right)} \frac{\delta^3 \sin^3(\omega t)}{6}}_{1\omega, 3\omega} + \underbrace{B \gamma^4 \cos(\gamma g) \frac{\delta^4 \sin^4(\omega t)}{24}}_{2\omega, 4\omega} , 
\end{eqnarray} 
where the frequency component of each term in this equation is highlighted. Expressing the intensity as $ I(t) = \sum \limits_{m}^{~} I_{m\omega} \sin{m \omega t}$, we can find expressions for the amplitude $I_{m\omega}$ of each frequency component:
 \begin{eqnarray}\label{eq:1om}
 I_{1\omega} =  - B \gamma \delta  \sin{\left(\gamma g \right)} + \frac{1}{8} B \delta^3  \gamma^3 \sin{\left( \gamma g \right)}, \\
 I_{2\omega}  =  \frac{1}{4} B \delta^2 \gamma^2  \cos{\left( \gamma g \right)} - \frac{1}{48} B\gamma^4 \delta^4 \cos(\gamma g), \\
 I_{3\omega} = - \frac{1}{24} B \delta^3   \gamma^3 \sin{\left( \gamma g \right)}, \label{eq:I3om}  \\
 I_{4\omega}  = \frac{1}{192} B \gamma^4 \delta^4 \cos(\gamma g).
 \end{eqnarray}
 In the main section of the paper these expressions were used to derive the expression for the ratio between $\frac{I_{3\omega}}{I_{1\omega}}$. This ratio cannot be determined in the case where $\sin{\left(\gamma g \right)}$ is close to zero. However, in that case the ratio $\frac{I_{4\omega}}{I_{2\omega}}$ can be used:
 \begin{equation}   
\frac{I_{4\omega}}{I_{2\omega}} = \frac{ \gamma^2 \delta^2}{48- 4 \gamma^2 \delta^2  },
\end{equation}
which gives for the amplitude:
\begin{equation}
\delta^* = \frac{4\sqrt{3{I_{4\omega}}/{I_{2\omega}}}}{\sqrt{4 \gamma^2 {I_{4\omega}}/{I_{2\omega}} + \gamma^2}}.
\end{equation}
The notation $\delta^*$ is used to denote that the ratio $\frac{I_{4\omega}}{I_{2\omega}}$ is used to calculate the amplitude, while $\delta$ implies that $\frac{I_{3\omega}}{I_{1\omega}}$ is used. Besides the two expressions that can be used for the amplitude, one can derive three more equations for the gap size $g$:
\begin{eqnarray}
   g_{32} = \frac{1}{\gamma}\left( \pi n + \arctan{\left(\frac{(12-\delta^2 \gamma^2)I_{3\omega}/I_{2\omega}}{-2 \delta \gamma}  \right)}\right),\\
 g_{41} = \frac{1}{\gamma}\left( \pi n + \arctan{\left(\frac{\delta^3 \gamma^3}{(24\delta^2 \gamma^2 - 192)I_{4\omega}/I_{1\omega}}\right)}\right),\\
 g_{43} = \frac{1}{\gamma}\left( \pi n + \arctan{\left(-\frac{\delta \gamma}{8I_{4\omega}/I_{3\omega}}\right)}\right). \label{eq:g43}
 \end{eqnarray}
 The equation for $g$ in the main section of the paper will be denoted by $g_{21}$. From here on, we will use $g^*$ to indicate that $\delta^*$ was used to calculate the average position, for example:
  \begin{eqnarray}
   g_{32}^* = \frac{1}{\gamma}\left( \pi n + \arctan{\left(\frac{(12-\delta^{*2} \gamma^2)I_{3\omega}/I_{2\omega}}{-2 \delta^* \gamma}  \right)}\right)
    \end{eqnarray}

\section{Disentangling optical and mechanical nonlinearities}
In the case of a Duffing nonlinearity, due to a cubic restoring force the assumption $x = \delta \sin(\omega t)$ will no longer be valid and we need to take into account the third harmonic term as well: $x = \delta \sin(\omega t) + \delta_{3\omega} \sin(3 \omega t)$. The resulting time-dependent intensity is given by:
\begin{eqnarray}
I = \underbrace{A+ B \cos{\left(\gamma g\right)}}_{DC}  \underbrace{- B \gamma \sin{\left(\gamma g \right)}\delta \sin(\omega t)}_{1\omega} \nonumber \\ \underbrace{ - B \gamma^2 \cos{\left( \gamma g \right)} \frac{\delta^2 \sin^2(\omega t)}{2}}_{2\omega} + \underbrace{B  \gamma^3 \sin{\left( \gamma g \right)} \frac{\delta^3 \sin^3(\omega t)}{6}}_{1\omega, 3\omega} + \underbrace{B \gamma^4 \cos(\gamma g) \frac{\delta^4 \sin^4(\omega t)}{24}}_{2\omega, 4\omega}  \nonumber \\ - \underbrace{B \gamma \sin(\gamma g) \delta_{3\omega} \sin(3 \omega t)}_{3\omega} - \underbrace{ B \gamma^2 \cos(\gamma g) \frac{\delta_{3\omega}^2 \sin^2 (3\omega t) }{2}}_{6\omega} + \underbrace{B \gamma^3 \sin(\gamma g) \frac{\delta_{3\omega}^3 \sin^3(3\omega t)}{6}}_{3\omega,9\omega}  \nonumber \\ +  \underbrace{B \gamma^4 \cos(\gamma g) \frac{\delta_{3\omega}^4\sin^4(3 \omega t)}{24} }_{6\omega,12\omega} 
\end{eqnarray} 
and the amplitudes up to the fourth harmonic are found by:
 \begin{eqnarray}\label{eq:1om}
 I_{1\omega} =  - B \gamma \delta  \sin{\left(\gamma g \right)} + \frac{1}{8} B \delta^3  \gamma^3 \sin{\left( \gamma g \right)},\\
 I_{2\omega}  =  \frac{1}{4} B \delta^2 \gamma^2  \cos{\left( \gamma g \right)} - \frac{1}{48} B\gamma^4 \delta^4 \cos(\gamma g), \\
 I_{3\omega} = - \frac{1}{24} B \delta^3   \gamma^3 \sin{\left( \gamma g \right)} - B\gamma\delta_{3\omega}\sin(\gamma g)-  \frac{1}{8} B \gamma^3 \delta_{3\omega}^3 \sin{(\gamma g)}, \label{eq:I3om}  \\
 I_{4\omega}  = \frac{1}{192} B \gamma^4 \delta^4 \cos(\gamma g).
 \end{eqnarray}
 From this extended analysis, we find that the ratio between $\frac{I_{4\omega}}{I_{2\omega}}$ is independent of the Duffing-type nonlinearity. 

 \begin{figure}
 \includegraphics{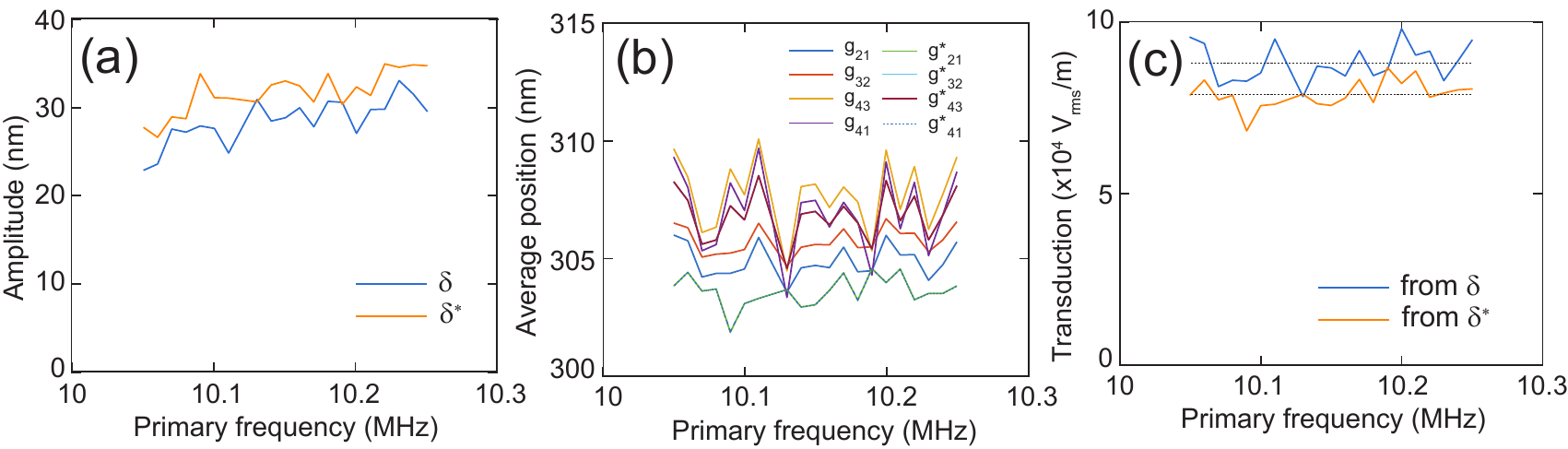}
 \caption{Result of the extended analysis. (a) amplitude obtained by using the ratio $I_{3\omega}/I_{1\omega}$, denoted by $\delta$ and the ratio $I_{4\omega}/I_{2\omega}$, denoted by $\delta^*$. A slightly higher amplitude is found for $\delta^*$, this is either due to the simplifying assumption behind eq. \ref{eq:cos}, or a small contribution due to mechanical nonlinearities. (b) Average position using eight different expressions, the notation $g^*$ is used to indicate that $\delta^*$ was used in to find the position. (c)  Transduction coefficient $\sqrt{\alpha}$, derived from both $\delta$ and $\delta^*$. \label{fig:s1}}
 \end{figure}
 \section{Experimental results}
Figure \ref{fig:s1} shows the results of the extended analysis. Using $\frac{I_{4\omega}}{I_{2\omega}}$ we find a slightly higher amplitude (Fig. \ref{fig:s1}(a)). This can be attributed to a contribution from the mechanical nonlinearities, but also to the error made by taking only the first term in the Fourier series in Eq. \ref{eq:cos}. The differences in the detected average postion all fall within 2\% of each other. The resulting transduction coefficient found by using $\frac{I_{4\omega}}{I_{2\omega}}$, is about 10\% lower then the coefficient found by using  $\frac{I_{3\omega}}{I_{1\omega}}$ (Fig. \ref{fig:s1}(c)). From the extended analysis we can therefore conclude that the upper bounds on the systemic errors in our system is 10\% on the transduction coefficient. For the average position, we find a systemic error of 2\%.

\end{document}